\documentclass[preprint,pre,10pt,twocolumn,footinbib,superscriptaddress,showpacs,showkeywords]{revtex4}
\usepackage{amsfonts}
\usepackage{amsmath}
\usepackage{amssymb}
\usepackage{graphicx}

\providecommand{\U}[1]{\protect\rule{.1in}{.1in}}

\begin{document}

\title[Extended canonical Monte Carlo methods]{Improving the efficiency of
Monte Carlo simulations \\
of systems that undergo temperature-driven phase transitions}
\author{L. Velazquez}
\affiliation{Departamento de F\'\i sica, Universidad Cat\'olica del Norte, Av. Angamos
0610, Antofagasta, Chile.}
\author{J.C. Castro-Palacio}
\affiliation{Department of Chemistry, University of Basel, Klingelbergstr. 80, 4056
Basel, Switzerland}
\keywords{Fluctuation theorems, Monte Carlo methods, Slow sampling problems}
\pacs{05.20.Gg, 02.70.Tt}
\date{\today}

\begin{abstract}
Recently, Velazquez and Curilef have proposed a methodology to extend Monte
Carlo algorithms based on canonical ensemble, which is aimed to overcome
slow sampling problems associated with temperature-driven discontinuous
phase transitions. We show in this work that Monte Carlo algorithms extended
with this methodology also exhibit a remarkable efficiency near a critical
point. Our study is performed for the particular case of 2D four-state Potts
model on the square lattice with periodic boundary conditions. This analysis
reveals that the extended version of Metropolis importance sample is more
efficient than the usual Swendsen-Wang and Wolff cluster algorithms. These
results demonstrate the effectiveness of this methodology to improve the
efficiency of MC simulations of systems that undergo any type of
temperature-driven phase transition.
\end{abstract}

\maketitle

\section{Introduction}

Many different algorithms have been proposed to overcome slow sampling
problems in large-scale Monte Carlo (MC) simulations. Most of them are based
on two types of strategies: (1) the substitution of local MC moves by a
simultaneous update of a large number of degrees of freedom, the so-called
\textit{cluster MC methods} \cite%
{SW,pottsm,Wolff,Edwards,Niedermayer,Evertz,Hasenbusch,Dress,Liu,MakSharma},
and (2) the use of histograms to extract information from MC simulations
combined with \emph{re-weighting techniques} to improve the statistics, such
as the multicanonical method and its variants \cite%
{BergM,Berg97,WangLandau,mc3,WangJStat}. Cluster MC methods are useful to overcome
slow sampling problems associated with a continuous Phase Transition (PT).
However, the application of nonlocal moves, by itself, does not help so much
in the presence of a discontinuous PT. For such cases, the consideration of
a re-weighting technique as the multicanonical method is more appropriate,
which reduces the size dependence of the decorrelation time from exponential
to a power-law behavior.

Not one of the above strategies seem to be sufficiently general to overcome
any type sampling problems of MC simulations. The success of clusters MC
algorithms is not universal because of the proper cluster moves seem to be
highly dependent on the system. In fact, efficient cluster MC methods have
only been found for a reduced number of models \cite%
{SW,pottsm,Wolff,Edwards,Niedermayer,Evertz,Hasenbusch,Dress,Liu,MakSharma}.
Multicanonical method and its variants have a general applicability.
However, the efficiency of these algorithms is not so significant to justify
their application to overcome slow sampling problems associated with
continuous PTs \cite{mc3}.

Recently, Velazquez and Curilef have introduced a different methodology to
overcome slow sampling problems associated with a temperature-driven
discontinuous PT \cite{vel-emc1,vel-emc2}. Their proposal is based on the
general equilibrium situation associated with fluctuation relation \cite%
{vel-efr,vel-tur}:
\begin{equation}
C=\beta ^{2}\left\langle \delta U^{2}\right\rangle +C\left\langle \delta
\beta _{\omega }\delta U\right\rangle ,  \label{fdr}
\end{equation}
which generalizes canonical relation \cite{Reichl,Landau}:
\begin{equation}  \label{can.fr}
C=\beta^{2}\left\langle \delta U^{2}\right\rangle
\end{equation}
between the heat capacity $C$ and the energy fluctuations \footnote{%
Boltzmann constant $k$ is assumed as the unity.}. This relation describes
the existence of a \emph{feedback perturbation} of the environment during
its thermodynamic interaction with the system. This mechanism is
characterized by the correlation function $\left\langle \delta \beta
_{\omega }\delta U\right\rangle $ between the system internal energy $U $
and the environmental inverse temperature $\beta _{\omega }=1/T^{\omega}$. A
relevant feature of fluctuation relation (\ref{fdr}) is its compatibility
with the existence of \textit{negative heat capacities} $C<0$ \cite%
{pad,Lyn3,gro1}.

The consideration of the above arguments in MC simulations enables a
considerable reduction of the dependence of the decorrelation times on
system size $N$, from exponential $\tau(N)\propto \exp(\gamma N)$ to a very
weak power-law behavior $\tau(N)\propto N^{\alpha}$ \footnote{%
Dependencies of decorrelation times considered in this work are expressed in
terms of system size (or the number of lattice sites) $N=L^{d}$ instead of
lattice linear size $L$. Consequently, dynamic critical exponents $\alpha$
are related to exponents $z$ of dependence $\tau(L)\propto L^{z}$ as $%
z=D*\alpha$, with $D$ being the lattice dimensionality.}. For example,
dynamic critical exponent $\alpha $ ranges from $0.14$ to $0.2$ in the case
of 2D seven and ten-state Potts model regardless one employees local or
nonlocal MC moves \cite{vel-emc1,vel-emc2}. Such an improvement is
significantly better than the one achieved applying the multicanonical
method and its variants to the same model systems, whose typical exponent $%
\alpha$ ranges from 2 to 2.5 \cite{BergM,Berg97}. We shall show in this work that
canonical MC algorithms extended with this methodology also exhibit a good
performance near a critical point. This claim is illustrated for the
particular case of 2D four-state Potts model \cite{mc3}. Our results
evidence the effectiveness of the present methodology to overcome slow
sampling problems associated with temperature-driven PT regardless its
continuous or discontinuous character.

\section{Overview of methodology}

\subsection{Theoretical backgrounds}

Methodology reviewed in this section is based on the consideration of \emph{%
generalized ensembles}. Many generalized ensembles that are employed in MC
simulations have no physical meaning, e.g.: multicanonical ensemble \cite%
{BergM,WangLandau,mc3,WangJStat}. However, this is non necessarily the case
of equilibrium situation considered in Ref.\cite{vel-efr} to derive
fluctuation relation (\ref{fdr}): a closed system composed of two systems A
and B with \textit{finite heat capacities} $C_{A}$ and $C_{B}$, which are
put in thermal contact among them and isolated of any external influence.
This situation can be implemented in a physical laboratory with an
acceptable accuracy. Moreover, this is the arrangement considered in
statistical mechanics to discuss thermal equilibrium conditions \cite%
{Reichl,Landau}. Curiously, standard textbooks of statistical mechanics
never refer to implications of equilibrium conditions concerning to states
with negative heat capacities \cite{pad,Lyn3,gro1}. For its own importance
in this work, let us start this section clarifying this question. We
recommend readers to see Ref.\cite{vel-efr} for further details.

As usual, the total energy $U_{T}$ and entropy $S_{T}$ of the closed system
are assumed as additive quantities, $U_{T}=U_{A}+U_{B}$ and $S=S_{A}+S_{B}$.
Maximization of entropy $S_{T}$ at constant energy $U_{T}$ demands the
stationary condition:
\begin{equation}
\frac{\partial S_{T}}{\partial U_{A}} =\frac{\partial S_{A}}{\partial U_{A}}-%
\frac{\partial S_{B}}{\partial U_{B}}=0\Rightarrow\frac{1}{T_{A}}=\frac {1}{%
T_{B}}\left(\equiv \beta\right),
\end{equation}
as well as the stability condition:
\begin{equation}  \label{stab}
\frac{\partial^{2}S_{T}}{\partial U_{A}^{2}}=\frac{\partial^{2} S_{A}}{%
\partial U^{2}_{A}}+\frac{\partial^{2} S_{B}}{\partial U^{2}_{B}}
<0\Rightarrow\frac{C_{A}C_{B} }{C_{A}+C_{B}}>0.
\end{equation}
We have considered here the microcanonical expressions:
\begin{equation}
\frac{1}{T_{\alpha}}=\frac{\partial S_{\alpha}}{\partial U_{\alpha}}\text{
and }\frac{\partial^{2}S_{\alpha}}{\partial U_{\alpha}^{2}}=-\frac{1}{%
T_{\alpha }^{2}C_{\alpha}},
\end{equation}
with $\alpha=\left( A,B\right) $. Accordingly, systems A and B can be found in
thermal equilibrium if they exhibit the same temperature and their heat
capacities satisfy one of the following stability conditions: (i) both
systems exhibit positive heat capacities, or (ii) a system exhibits a
negative heat capacity, e.g.: $C_{A}<0$, and the other a positive heat
capacity $C_{B}>0$ that satisfies the following inequality:
\begin{equation}
C_{B}<\left\vert C_{A}\right\vert .  \label{Thirring}
\end{equation}

Condition (\ref{Thirring}) was obtained by Thirring almost forty years ago
\cite{Thirring}. Accordingly, a system with negative heat capacity cannot be
found in thermal equilibrium with an environment that exhibits an infinite
heat capacity. e.g.: under thermodynamic influence of the natural
environment. In other words, canonical ensemble is unable to study systems
with negative heat capacities. However, these systems can be analyzed
considering the thermal contact with an environment that exhibits \emph{a
finite heat capacity}. This conclusion is specially relevant for MC
simulations. An unexpected consequence of the above analysis concerns to
so-called \emph{zeroth law of thermodynamics} \cite{Guggenheim}, which
states: if two systems are both in thermal equilibrium with a third system
then they are in thermal equilibrium with each other. Although this \emph{law%
} helps to define the notion of temperature, its validity is restricted to
systems exhibiting positive heat capacities. Two identical systems, that are
initially prepared in the same macroscopic state, cannot be in thermal
equilibrium if such macroscopic state exhibits a negative heat capacity.
Noteworthy that each system can remain in thermal equilibrium with a third
system exhibiting a positive heat capacity, whenever its obeys Thirring
inequality (\ref{Thirring}). This violation of zeroth law of thermodynamics
was recently discussed in the literature \cite{Ramirez}.

Let us now regard system B as an \emph{environment} in order to study
thermodynamic properties of system A. Since heat capacity of this
environment is finite, its temperature $T_{B}$ will be affected by the
energy interchange with the system A. Considering $\delta T_{B}=-\delta
U_{A}/C_{B}\Rightarrow\delta\beta_{B}=\beta^{2}\delta U_{A}/C_{A}$, one
obtains from (\ref{fdr}) the following result:
\begin{equation}  \label{fluct.deriv}
\frac{C_{A}C_{B}}{C_{A}+C_{B}}=\beta^{2}\left\langle \delta
U_{A}^{2}\right\rangle .
\end{equation}
Stability condition (\ref{stab}) is derived from positivity of r.h.s. of Eq.(%
\ref{fluct.deriv}), but this time from a \textit{fluctuational viewpoint}.
Relation (\ref{fluct.deriv}) drops to canonical fluctuation relation (\ref%
{can.fr}) in the limit $C_{B}\rightarrow+\infty$, as well as the
microcanonical result $\left\langle \delta U_{A}^{2}\right\rangle
\rightarrow0$ when $C_{B}\rightarrow0^{+}$. Accordingly, this type of
equilibrium situation can be associated with a family of \textit{generalized
ensembles} that contains microcanonical and canonical ensembles as
particular cases. As expected, the exact mathematical form of each ensemble
depends on the system B acting as environment.

Phenomenon of negative heat capacity has been regarded as an anomalous
behavior. Precisely, their existence is incompatible with results of
classical fluctuation theory \cite{Reichl,Landau}, e.g.: fluctuation
relation (\ref{can.fr}). However, this incompatibility arises because of the
restricted applicability of some conventional assumptions. Specifically, the
macroscopic state of the environment can be affected by influence of the
system under study. Such an \emph{environmental feedback perturbation} is
systematically omitted when one employs traditional ensembles such as
Boltzmann-Gibbs distributions \cite{Reichl,Landau}:
\begin{equation}
\omega_{BG}\left( U,X|T,Y\right) =\frac{1}{Z}\exp\left[ -\beta\left(
U+XY\right) \right] ,
\end{equation}
where the environmental inverse temperature $\beta=1/T$ and the external
generalized forces $Y$ (e.g.: pressure $p$, magnetic and electric fields $%
\mathbf{H}$ and $\mathbf{E}$, etc.) are assumed as \textit{constant control
parameters} for the energy $U$ and the generalized displacements $X$ (e.g.:
volume $V$, magnetization $\mathbf{M}$ and electric polarization $\mathbf{P}$%
, etc.). As expected, more general equilibrium situations involve
non-vanishing correlations such as $\left\langle \delta\beta\delta
U\right\rangle$ or $\left\langle \delta Y\delta X\right\rangle$. Fluctuation
theorems associated with these equilibrium situations provide a suitable
treatment for states with negative heat capacities as well as other
anomalies in response functions \cite{vel-geft}.

\subsection{Application to MC simulations}

The inclusion of a feedback effect $\left\langle \delta \beta_{\omega
}\delta U\right\rangle $ to extend any canonical MC algorithm is achieved
replacing the constant inverse temperature $\beta $ of the canonical
ensemble by an \textit{effective inverse temperature} $\beta _{\omega
}\left( U\right) $:
\begin{equation}  \label{beta.omega}
\beta _{\omega }\left( U\right)=-\frac{\partial}{\partial U}\log\omega(U),
\end{equation}
which depends on the energy $U$ of system under study. This effective
inverse temperature corresponds to an environmental influence whose
probability weight $\omega(U)$ differs from the one associated with
canonical ensemble:
\begin{equation}
\omega(U)\not=\omega_{c}(U|\beta)=\frac{1}{Z(\beta)}\exp\left(-\beta
U\right).
\end{equation}

This type of arguments were employed by Gerling and H\"{u}ller to propose
\emph{dynamic ensemble} MC method \cite{Gerling}. These authors considered
as environment an ideal gas with $N$ degrees of freedom. Analysis of
detailed balance led them to introduce an effective inverse temperature $%
\beta_{\omega}=(N-2)/2Nk_{b}$. Here, $k_{b}=(U_{T}-U)/N$ is the mean kinetic
energy per particle for the ideal gas, while $U_{T}$ is the total energy.
Effective inverse temperature $\beta_{\omega}$ is \emph{adjusted dynamically}
during the course of MC simulation. Data are then obtained by computing the
mean value of the energy $\left\langle U\right\rangle$ and the mean value of
the temperature from $2\left\langle k_{b}\right\rangle$. This method allows
to detect the presence of states with negative heat capacities whenever
Thirring inequality (\ref{Thirring}) is fulfilled.

Methodology proposed by Velazquez and Curilef includes three important
modifications for dynamic ensemble MC method of Gerling and H\"{u}ller \cite%
{vel-emc1,vel-emc2}: (i) the consideration of a more suitable \textit{%
generalized ensemble}, (ii) the employment of a \emph{point statistical
estimation} to obtain the relevant microcanonical dependencies and reduce
incidence of finite size effects, and finally, (iii) the \emph{optimization
of efficiency} considering a more active control on the system fluctuating
behavior. For the rest of this section, let us detailedly explain these
modifications.

\subsection{Gaussian ensemble and its implementation}

Let us consider the power-expansion of environmental inverse temperature $%
\beta_{\omega}(U)$ around the mean value of the energy $U_{e}=\left\langle U
\right\rangle$:
\begin{equation}  \label{expansion}
\beta_{\omega}(U)=\beta+\sum^{\infty}_{n=1}a_{n}(U-U_{e})^{n}.
\end{equation}
Assuming that energy fluctuations $\delta U=U-U_{e}$ are sufficiently small,
power-expansion (\ref{expansion}) can be restricted to first-order
approximation, $\beta_{\omega}=\beta+\lambda\delta U/N$. Coupling constant $%
\lambda=Na_{1}$ plays a role of control parameter in conjunction with the
expectation value of inverse temperature $\beta=\left\langle
\beta_{\omega}\right\rangle $. Substituting this ansatz into Eq.(\ref{fdr}),
one obtains the fluctuation relations:
\begin{equation}
\left\langle \delta U^{2}\right\rangle =\frac{N}{\beta^{2}N/C+\lambda }%
\mbox{ and }\left\langle \delta\beta_{\omega}^{2}\right\rangle =\frac{1}{N}%
\frac{\lambda^{2}}{\beta^{2}N/C+\lambda}  \label{dispersions}
\end{equation}
as well as the stability condition:
\begin{equation}
\beta^{2}N/C+\lambda>0.  \label{stab.cond}
\end{equation}
Noteworthy that the size dependencies of the energy $\Delta U$ and inverse
temperature $\Delta \beta_{\omega}$ dispersions ($\Delta x\equiv\sqrt{
\left\langle \delta x^{2}\right\rangle }$) behave as $\Delta U\propto\sqrt{N}
$ and $\Delta\beta_{\omega}\propto1/\sqrt{N}$ for a short-range interacting
system. As expected, linear approximation $\beta_{\omega}=\beta+\lambda%
\delta U/N$ is good as long as the system size $N$ be sufficiently large. If
coupling constant $\lambda$ obeys the stability condition (\ref{stab.cond}),
statistical ensemble associated with the present equilibrium situation
becomes equivalent to the microcanonical ensemble in the thermodynamic
limit:
\begin{equation}
\lim_{N\rightarrow\infty}\frac{\Delta U}{U}=\lim_{N\rightarrow\infty
}\Delta\beta_{\omega}=0  \label{equivalence}
\end{equation}
\textit{regardless} the positive or negative character of heat capacity $C$
of the system under study. The origin of the exponential dependence with $N$
of the decorrelation time in MC simulations $\tau(N)\propto\exp( \gamma N)$
is due to the \textit{multimodal character} of the energy distribution
function within the canonical ensemble \cite{mc3}. Such a bimodal character
of energy distributions is associated with the existence of macrostates with
negative heat capacities. Since the ensemble equivalence ensures the
existence of only one peak, MC simulations based on the present equilibrium
situation cannot undergo this type of slow sampling problem.

Equilibrium situation previously described is implemented assuming a linear
dependence of the environmental inverse temperature on the system energy, $%
\beta_{\omega}(U)= \beta_{s}+\lambda_{s}(U-U_{s})/N$, with $(U_{s},
\beta_{s},\lambda_{s})$ being three seed parameters, where $U_{s}$ and $%
\beta_{s}$ are roughly estimates of the expectation values $\left\langle
U_{\omega}\right\rangle $ and $\left\langle \beta_{\omega}\right\rangle $.
According to Eq.(\ref{beta.omega}), this choice corresponds to the \textit{%
gaussian ensemble} \cite{Challa,Hetherington}:
\begin{equation}
\omega_{G}\left( U\right) =\frac{1}{Z_{\lambda}\left( \beta_{s}\right) }\exp%
\left[ -\beta_{s} U-\frac{1}{2N}\lambda_{s}\left( U-U_{s}\right) ^{2}\right]
\label{GEns}
\end{equation}
with parameter $\lambda_{s}\geq 0$. Gaussian ensemble describes intermediate
equilibrium situations between the usual thermal contact (canonical
ensemble) when $\lambda_{s}\rightarrow 0^{+}$ and energy isolation
(microcanonical ensemble) when $\lambda_{s}\rightarrow +\infty$. The bath
associated with this ensemble corresponds to a hypothetical substance whose
heat capacity decreases with temperature as $C_{B}\propto 1/T^{2}$ \footnote{%
Dependence of the form $C\propto 1/T^{2}$ is observed in the hight
temperature limit of a paramagnetic ideal gas.}. Gaussian ensemble (\ref%
{GEns}) provides several advantages to improve canonical MC simulations. In
particular, its mathematical form makes more easy the analysis of detailed
balance and the point statistical estimation.

Let $W^{c}\left(U_{i}\rightarrow U_{j};\beta\right)$ be the transition
probability of a given canonical MC algorithm, which satisfies detailed
balance condition:
\begin{equation}
\frac{W^{c}\left( U_{i}\rightarrow U_{j};\beta\right)}{W^{c}\left(
U_{j}\rightarrow U_{i};\beta\right)}=\exp\left(-\beta\delta U_{ij}\right),
\end{equation}
where $\delta U_{ij}=U_{j}-U_{i}$ is energy change after transition. The
detailed balance condition corresponding to gaussian ensemble (\ref{GEns})
can be satisfied considering the following transition probability $W\left(
U_{i}\rightarrow U_{j}\right)$:
\begin{equation}
W\left( U_{i}\rightarrow U_{j}\right)=W^{c}\left( U_{i}\rightarrow U_{j};
\beta^{t}_{\omega}\right),
\end{equation}
where $\beta_{\omega}^{t}=\left(
\beta_{\omega}^{i}+\beta_{\omega}^{j}\right)/2$ is the \textit{transition
inverse temperature} \cite{vel-emc2}, with $\beta_{\omega}^{i}$ and $%
\beta_{\omega}^{j}$ being the environmental inverse temperatures at the
initial and the final configurations respectively, $\beta_{\omega}^{i}=%
\beta_{\omega}\left(U_{i}\right) $ and $\beta_{\omega
}^{j}=\beta_{\omega}\left( U_{j}\right)$. This result follows from the
identity:
\begin{equation}
\frac{W\left( U_{i}\rightarrow U_{j}\right)}{W\left( U_{j}\rightarrow
U_{i}\right)}=\frac{\omega_{G}\left( U_{j}\right)}{\omega_{G}\left(
U_{i}\right)}\equiv\exp\left(-\beta^{t}_{\omega}\delta U_{ij}\right),
\end{equation}
which is obtained from the mathematical form of gaussian ensemble.
Accordingly, one should replace the constant inverse temperature $\beta$ of
any canonical MC algorithm by the transition inverse temperature $%
\beta_{\omega}^{t}$. Unfortunately, the application of this result demands
to know, \textit{a priori}, the final configuration $X_{j}$ of the system
with energy $U_{j}$. Therefore, this method can only be applied to extend
local MC algorithms such as Metropolis importance sample \cite%
{metro,Hastings} or Glauber dynamics \cite{Glauber}. The extending of
cluster canonical MC algorithms is also possible, but their implementation
is carried out dividing each MC moves into two steps \cite{vel-emc2}:

\begin{enumerate}
\item To obtain a virtual configuration $X_{j}$ with energy $U_{j}$ through
a canonical cluster MC method using the inverse temperature $%
\beta^{i}_{\omega}$ of the initial configuration $X_{i}$ with energy $U_{i}$;

\item To accept the virtual configuration $X_{j}$ using the acceptance
probability $w_{i\rightarrow j}$:
\begin{equation}
w_{i\rightarrow j}=\min\left\{ 1,\frac{W^{j}_{j\rightarrow i}}{%
W^{i}_{i\rightarrow j}}\exp\left( -\beta_{\omega}^{t}\delta U_{ij}\right)
\right\}.  \label{wif}
\end{equation}
\end{enumerate}

The terms $W^{i}_{i\rightarrow j}=W^{c}\left[ U_{i}\rightarrow
U_{j};\beta_{\omega}^{i}\right]$ and $W^{j}_{j\rightarrow i}=W^{c}\left[
U_{j}\rightarrow U_{i};\beta_{\omega}^{j}\right]$ represent the transition
probabilities of the direct and the reverse process, respectively. Thus, the
transition probability of the global process can be expressed as:
\begin{equation}
W\left( U_{i}\rightarrow U_{j}\right)=W^{c}\left( U_{i}\rightarrow
U_{j};\beta^{i}_{\omega}\right)w_{i\rightarrow j}.
\end{equation}
In general, values of the acceptance probability $w_{i\rightarrow j}$ are
close to the unity because of the change of the inverse temperature $%
\delta\beta^{ij}_{\omega}=\beta^{j}_{\omega}-\beta^{i}_{\omega}$ and the
energy change $\delta U_{ij}$ are very small if the system size $N$ is
sufficiently large.

\subsection{Point statistical estimation}

\label{PSE}

By definition, statistical expectation values of macroscopic observables are
\emph{ensemble-dependent}, that is, they depend on the concrete equilibrium
situation associated with a given statistical ensemble. To avoid this
arbitrariness, one should perform the calculation of microcanonical quantities
derived from the system entropy $S(U)$, such as the microcanonical caloric
curve $\beta(U)=\partial S(U)/\partial U$ and the curvature curve $%
\kappa(U)=-N\partial ^{2}S(U)/\partial U^{2}$. Notice that this second
quantity is directly related to the microcanonical heat capacity $C$ as $%
\kappa =\beta ^{2}N/C$.

In multicanonical algorithms and other re-weighting techniques, the
microcanonical dependencies $\beta(U)$ and $\kappa(U)$ can be obtained by
direct numerical differentiation of the entropy $S(U)$, which was previously
estimated using energy histograms. However, this procedure increases the
statistical errors associated with any MC calculations, whose incidence is
more significant with a larger order of differentiation \cite{vel-emc2}. A
more precise calculation is performed using the \textit{point statistical
estimation} at the equilibrium energy $U_{e}$, which is related to the
thermal equilibrium condition $\beta _{\omega }\left( U_{e}\right)=\beta
\left( U_{e}\right) =\beta _{e}$. The estimation of microcanonical
quantities $(U_{e},\beta _{e},\kappa _{e})$ is based on the asymptotic
tendency of the energy distribution to adopt a \emph{gaussian form} in the
thermodynamic limit $N\rightarrow+\infty$. In analogous way to dynamic
ensemble MC method \cite{Gerling}, \emph{estimation of microcanonical
dependencies is only exact in the thermodynamic limit}. However, the
incidence of finite size effects is considerably reduced using the following
expressions \cite{vel-emc2}:
\begin{eqnarray}
U_{e} =\left\langle U\right\rangle -\frac{1-\psi _{1}}{2\left\langle \delta
U^{2}\right\rangle }\left\langle \delta U^{3}\right\rangle +O\left( \frac{1}{%
N^{3}}\right) ,  \notag \\
\beta _{e} =\left\langle \beta _{\omega }\right\rangle -\lambda \frac{1-\psi
_{1}}{2N\left\langle \delta U^{2}\right\rangle }\left\langle \delta
U^{3}\right\rangle +O\left( \frac{1}{N^{3}}\right) ,  \label{EE} \\
\kappa _{e} =\frac{1-\psi _{1}-\lambda \left\langle \delta
U^{2}\right\rangle /N}{\left\langle \delta U^{2}\right\rangle /N}+O\left(
\frac{1}{N^{2}}\right) ,  \notag
\end{eqnarray}
where $\psi _{1}=\frac{6}{5}\epsilon _{2}+\frac{11}{30}\epsilon _{1}$ is a
second-order correction term defined from the cumulants $\epsilon _{1}$ and $%
\epsilon _{2}$:
\begin{equation}
\epsilon _{1}=\frac{\left\langle \delta U^{3}\right\rangle ^{2}}{%
\left\langle \delta U^{2}\right\rangle ^{3}},~\epsilon _{2}=1-\frac{%
\left\langle \delta U^{4}\right\rangle }{3\left\langle \delta
U^{2}\right\rangle ^{2}}.  \label{cumulants}
\end{equation}
Accordingly, one should proceed the MC calculation of the statistical
expectation values $\left\langle U\right\rangle$ and $\left\langle \beta
_{\omega }\right\rangle$, as well as the $n$-moments of energy $\left\langle
\delta U^{n}\right\rangle$ with $n=(2, 3, 4)$. Noteworthy that expression
for curvature $\kappa_{e}=\beta^{2}_{e}N/C_{e}$ represents a second-order
improvement of the energy fluctuations considered in Eq.(\ref{dispersions}).
This same calculations enable us to obtain a roughly estimation for the
third and the four-order derivatives of the entropy:
\begin{eqnarray}
\zeta^{3}_{e}&=&N^{2}\frac{\partial^{3}S(U_{e})}{\partial U^{3}}=N^{2}\frac{%
\left\langle \delta U^{3}\right\rangle}{\left\langle \delta
U^{2}\right\rangle ^{3}}\left(1-3\psi_{1}\right) +O\left( \frac{1}{N^{2}}%
\right) ,  \notag \\
\zeta^{4}_{e}&=&N^{3}\frac{\partial^{4}S(U_{e})}{\partial U^{4}}=-\psi_{2}%
\frac{N^{3}}{\left\langle \delta U^{2}\right\rangle ^{3}}+O\left( \frac{1}{N}%
\right),
\end{eqnarray}
where $\psi_{2}=\frac{12}{5}\epsilon _{2}+\frac{41}{15}\epsilon _{1}$.
Derivation of the above formulae was discussed in appendix section of Ref.%
\cite{vel-emc2}. The same ones were obtained for the particular case of
gaussian ensemble (\ref{GEns}), and their applicability is associated to
licitness of gaussian approximation for describing system fluctuating
behavior. This means that the seed parameters $(U_{s},\beta_{s},\lambda_{s})$
of gaussian ensemble (\ref{GEns}) should be carefully chosen to guarantee
applicability of gaussian approximation. The way to achieve this goal will
be explained at the end of the next subsection.

\subsection{Efficiency factor and its optimization}

The efficiency of MC methods is commonly characterized by the \textit{%
decorrelation time} $\tau $, that is, the minimum number of MC steps needed
to generate effectively independent, identically distributed samples in the
Markov chain \cite{mc3}. This quantity will be calculated as follows:
\begin{equation}
\tau =\lim_{k\rightarrow \infty }\tau _{k}=\lim_{k\rightarrow \infty }\frac{%
k\cdot var\left( u_{k}\right) }{var\left( u _{1}\right) },
\label{decorrelation}
\end{equation}%
where $var\left( u _{k}\right) =\left\langle u_{k}^{2}\right\rangle
-\left\langle u _{k}\right\rangle ^{2}$ is the variance of $u _{k}$, which
is defined as the arithmetic mean of the energy per particle $u=U/N $ over $%
k $ samples (consecutive MC steps):
\begin{equation}
u _{k}=\frac{1}{k}\sum_{i=1}^{k}u _{i}.
\end{equation}

However, the decorrelation time $\tau $ provides a partial characterization
about the efficiency in the case of the extended canonical MC methods. To
clarify this idea, let us consider the number of MC steps $S$ needed to
obtain a point of the caloric curve $\beta(u)$ with a precision $\delta
u^{2}+\delta\beta ^{2}<a^{2} $. This quantity can be estimated in terms of
the total dispersion $\Delta_{T}^{2}=\left\langle \delta
U^{2}\right\rangle/N+N\left\langle \delta\beta_{\omega}^{2}\right\rangle$
and the decorrelation time $\tau $ as follows:
\begin{equation}  \label{num.steps}
S=\tau \Delta _{T}^{2}/Na^{2}.
\end{equation}
The total dispersion $\Delta _{T}^{2}$ is kept fixed for canonical ensemble,
and hence, a canonical MC algorithm is more efficient as smaller is its
decorrelation time $\tau$. However, the total dispersion $\Delta _{T}^{2}$
is \emph{ensemble-dependent}, e.g.: this quantity depends on the control
parameters $(U_{s},\beta_{s},\lambda_{s})$ of gaussian ensemble (\ref{GEns}%
). According to expression (\ref{num.steps}), an extended canonical MC
algorithm is more efficient as smaller is its \textit{efficiency factor}:
\begin{equation}  \label{eta}
\eta=\tau \Delta _{T}^{2}.
\end{equation}
The efficiency factor (\ref{eta}) depends on both decorrelation time $\tau$
and the system fluctuating behavior. Moreover, decorrelation time $\tau$
depends on both the statistical ensemble and the concrete canonical MC
algorithm. The explicit mathematical form of the decorrelation time $\tau$
in terms of control parameters of a given MC calculation is difficult to
precise. The simplest criterion to reduce the efficiency factor $\eta$ is
minimizing the total dispersion $\Delta_{T}^{2}$, that is, by introducing a
more active control on the system fluctuating behavior \cite%
{vel-emc1,vel-emc2}. Using the expressions of equation (\ref{dispersions}),
the lower-bound of the total dispersion $\Delta_{T}^{2}$ and the optimal
value of the control parameter $\lambda_{s}$ are the following:
\begin{equation}
\lambda_{s}=\lambda_{\Delta}\left( \kappa_{e}\right) =\sqrt {1+\kappa_{e}^{2}%
}-\kappa_{e}\mbox{ and }\min(\Delta_{T}^{2})=2\lambda_{\Delta },
\end{equation}
where $\kappa_{e}$ is the curvature at the energy point $U_{e}$.
Accordingly, the optimal value for the parameter $\lambda_{s}$ demands to
consider a roughly estimation of the curvature $\kappa_{e}$.

Seeds parameters $(U_{s},\beta_{s},\lambda_{s})$ for a given MC run can be
specified using the microcanonical estimates $(U_{e},\beta_{e},\kappa_{e})$
obtained from a previous simulation run. We have employed in this work the
following iterative scheme:
\begin{equation}
U^{j+1}_{s}=U^{j}_{e}+
\varepsilon;\,\beta^{j+1}_{s}=\beta^{j}_{e}-\kappa^{j}_{e}\varepsilon%
\mbox{
and }\lambda^{j+1}_{s}=\lambda_{\Delta}(\kappa^{j}_{e}),
\end{equation}
with $\varepsilon$ being a small energy step. Noteworthy that scheme for $%
\beta^{j+1}_{s}$ is simply first-order power-expansion of microcanonical
inverse temperature, $\beta^{j+1}_{s}=\beta(U^{j}_{e}+\varepsilon)=%
\beta(U^{j}_{e})+\beta^{\prime}(U^{j}_{e})\varepsilon+O(\varepsilon^{2})$.
Moreover, we have assumed a zero-order approximation for curvature $%
\kappa^{j+1}_{e}=\kappa(U^{j}_{e}+\varepsilon)=\kappa^{j}_{e}+O(\varepsilon)$%
. The initial values of the seed parameters $(U_{s},\beta_{s},\lambda_{s})$
could be estimated from any canonical MC algorithm far from the region of
temperature-driven PT. Sometimes, it is recommendable to consider a \emph{%
variable energy step} $\varepsilon$, overall, in those energy regions where
the absolute values of microcanonical curvature curve $\kappa(U)$ are
sufficiently large. We have employed in this work the following rule $%
\varepsilon=\varepsilon_{0}/\sqrt{1+\kappa^{2}_{e}}$, with $\varepsilon_{0}$
being the energy step near critical point where $\kappa_{e}\simeq 0$. Notice
that this rule guarantees, approximately, \emph{a constant arc-length}
between neighboring points of microcanonical caloric curve $\beta$ versus $U$%
. This feature can be checked in FIG.\ref{microestimates.eps}.

\section{Efficiency near a critical point}

\subsection{Temperature-driven continuous PT}

Ensemble equivalence is always ensured in the case of a temperature
driven-continuous PT. Slow sampling problems in systems that undergo this
type of PT are consequence of the large increasing of the energy
fluctuations and the heat capacity $C$ when the inverse temperature $\beta$
of canonical ensemble approaches the critical point $\beta_{c}$. As
discussed elsewhere \cite{Reichl}, the fluctuating behavior observed here
can be associated with the existence of \textit{large correlation length} $%
\xi$ among the system constituents.

The incidence of slow sampling problems could be significantly reduced if
such strong correlations could be \textit{avoided by some external influence}%
. If possible, the relaxation times of averages of physical observables
could be good enough even using local MC moves. Such a reduction of
correlation length $\xi$ can also be achieved considering the feedback
perturbation of the environment. According to Eqs.(\ref{dispersions}) and (%
\ref{stab.cond}), the coupling constant $\lambda $ acts as a control
parameter of the system thermodynamic stability and fluctuating behavior.
Canonical fluctuation relation (\ref{can.fr}) predicts that the energy
dispersion $\Delta U\rightarrow\infty$ when $C\rightarrow\infty$. However,
\textit{the quantity $\Delta U$ remains finite whenever the stability
condition (\ref{stab.cond}) applied}, that is, if the coupling constant $%
\lambda>0$ when $C\longrightarrow\infty$. Since the energy fluctuations are
kept finite at the critical point, the underlying correlation length $\xi$
among the system constituents should be reduced.

\begin{figure*}[tbp]
\begin{center}
\includegraphics[width=7.0in ]{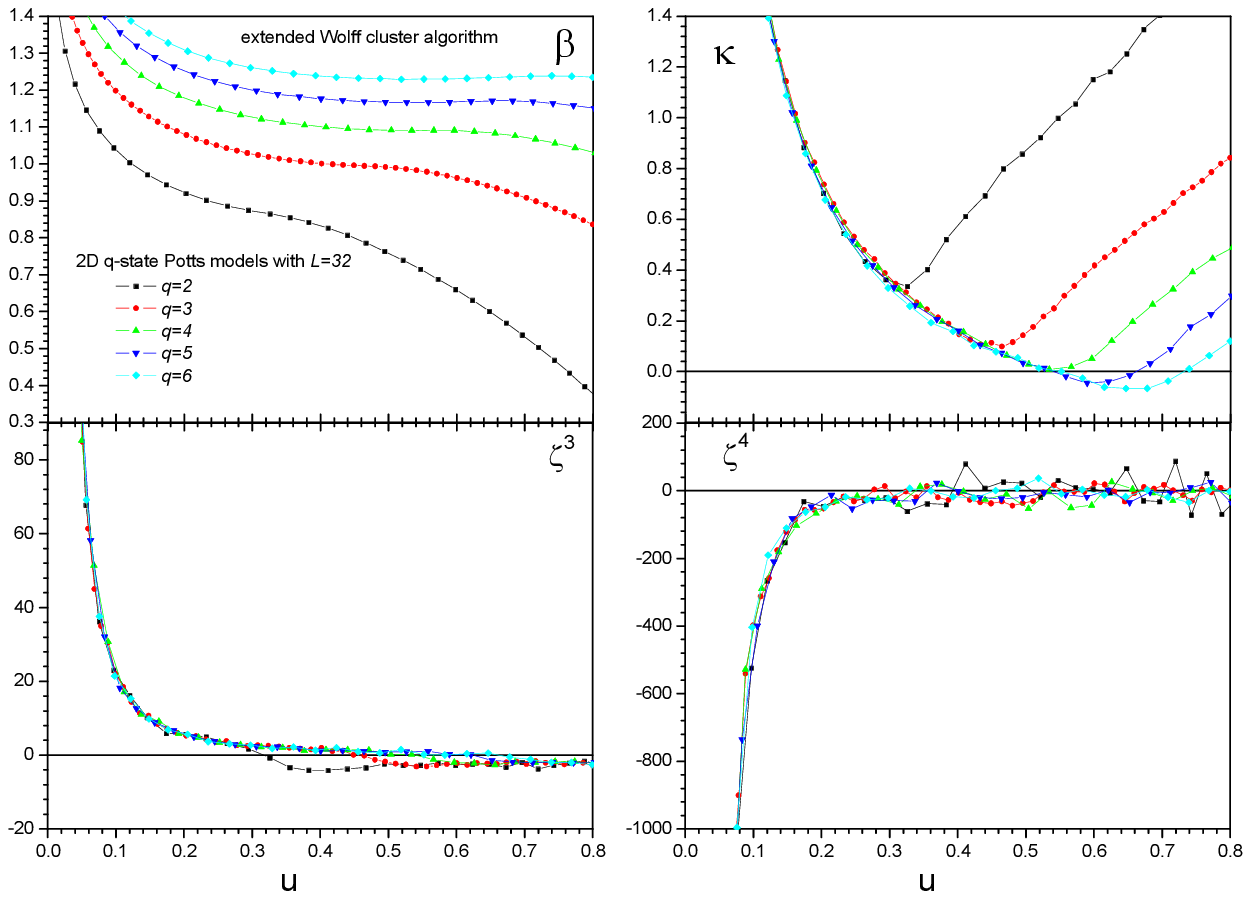}
\end{center}
\caption{(Color online). Microcanonical estimates of 2D $q$-state Potts model for $q=2-6$
with $L=32$, which were obtained from extended version of Wolff cluster
algorithm: the inverse temperature $\protect\beta(u)=\partial s(u)/\partial
u $, the curvature curve $\protect\kappa(u)=-\partial^{2} s(u)/\partial u^{2}
$, and third and four partial derivatives, $\protect\zeta^{3}(u)=%
\partial^{3} s(u)/\partial u^{3}$ and $\protect\zeta^{4}(u)=\partial^{4}
s(u)/\partial u^{4}$, with $s(u)$ and $u$ being the entropy and the energy
per site, respectively.}
\label{microestimates.eps}
\end{figure*}

\subsection{Potts model and its MC algorithms}

For the sake of convenience, let us consider the $q$-state Potts model \cite%
{mc3}:
\begin{equation}  \label{potts}
H=-\sum_{(i,j)}\delta_{\sigma_{i}\sigma_{j}}
\end{equation}
defined over a square lattice $L\times L$ with periodic boundary conditions,
where $\sigma_{i}=(1,2,\ldots q)$ is the spin variable of the $i$-th site
and the sum in (\ref{potts}) involves all nearest-neighbors. This family of
toy models undergoes both continuous and discontinuous PT at $%
\beta_{c}=\ln\left(1+\sqrt{q}\right)$ in the thermodynamic limit $%
L\rightarrow\infty$. Their MC study can be performed using different
canonical MC algorithms. Specifically, we will consider Metropolis
importance sample \cite{metro} as a local MC method, as well as
Swendsen-Wang and Wolff cluster algorithms \cite{SW,pottsm,Wolff} as
examples of nonlocal MC methods. These cluster MC methods are easily
extended with the application of the present methodology. Firstly, we need
to obtain the transition probabilities $W^{c}_{i\rightarrow j}$ and $%
W^{c}_{j\rightarrow i}$ that appear in the acceptance probability (\ref{wif}%
). Denoting by $p_{i}=1-e^{-\beta _{\omega }^{i}}$ and $p_{j}=1-e^{-\beta
_{\omega }^{j}}$ the acceptance probabilities of bonds for the direct and
reverse processes, the transition probabilities $W^{i}_{i\rightarrow j}$ and
$W^{j}_{j\rightarrow i}$ are expressed as follows:
\begin{equation}
W^{i}_{i\rightarrow j}=p_{i}^{b_{a}}\left( 1-p_{i}\right)
^{b_{p}+b_{d}},W^{j}_{j\rightarrow i}=p_{j}^{b_{a}}\left( 1-p_{j}\right)
^{b_{p}+b_{c}}.
\end{equation}
Here, $b_{a}$ and $b_{p}+b_{d}$ are the numbers of inspected bonds which
have been accepted and rejected in the direct process, respectively.
Moreover, $b_{d}$ is the number of rejected bonds which have been destroyed
in the final configuration $X_{f}$, while $b_{c}$ is number of created
bonds. Notice that the energy change $\delta U_{if}=b_{d}-b_{c}$. The
integer numbers $(b_{a},b_{d}, b_{c}, b_{p})$ should be obtained for each
cluster move.

\begin{figure}[tbp]
\begin{center}
\includegraphics[
width=3.5in ]{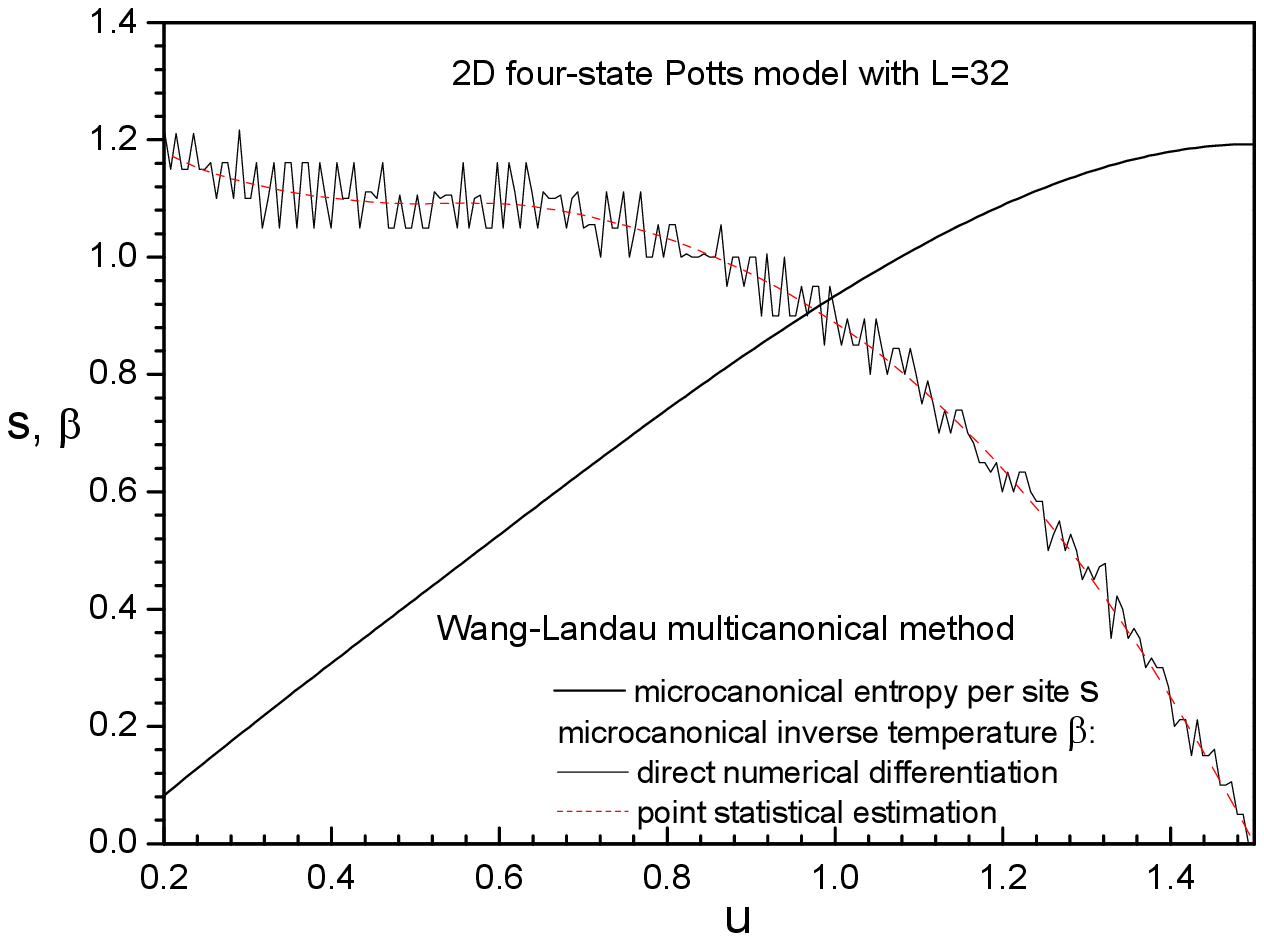}
\end{center}
\caption{(Color online). Entropy per site $s(u)$ and microcanonical inverse temperature $%
\protect\beta(u)$ of 2D four-state Potts model estimated from Wang-Landau
multicanonical algorithm. Here, the variable $u$ denotes the energy per
site, $u=U/N$, with $N=L^{2}$.}
\label{wanglandau.eps}
\end{figure}

\begin{figure}[tbp]
\begin{center}
\includegraphics[width=3.5in ]{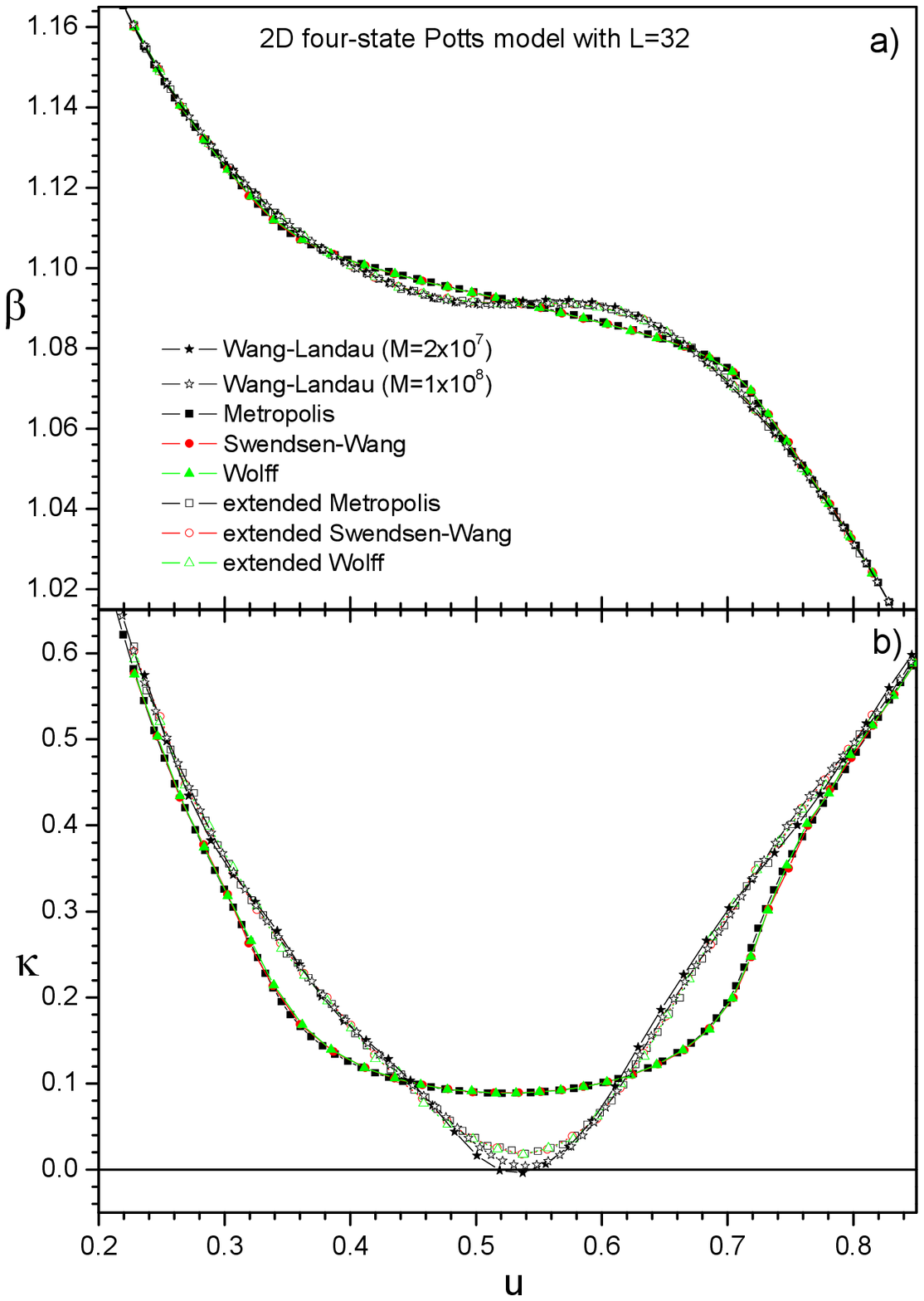}
\end{center}
\caption{(Color online). Energy dependence of inverse temperature $\beta$ and curvature $\kappa$ obtained from three different canonical MC algorithms and their extended versions in the case of 2D four-state Potts model with $L=32$. Each point estimated with these MC algorithms was obtained from a simulation run with a number of steps $M=4\times10^{4}\tau$, with $\tau$ being decorrelation time of this simulation run. Results of Wang-Landau multicanonical method are employed here as reference, which were obtained from two different simulation runs with $M=2\times10^{7}$ and $M=1.1\times10^{8}$ steps.}
\label{thermo.comp.eps}
\end{figure}

\subsection{Numerical simulations}

We have shown in FIG.\ref{microestimates.eps} several microcanonical
dependencies of $q$-state Potts model with $L=32$ and $q=2-6$, which were
estimated using the extended version of Wolff cluster algorithm and the
point statistical estimation (\ref{EE}). Each point of these curves was
obtained considering $M=4\times10^{4}\tau$ iterations for each MC run, with $%
\tau$ being its associated decorrelation time. The convergence of four-order
derivative $\zeta^{4}(u)$ is less significant than the other microcanonical
dependencies. However, this is a reasonable result taking into consideration
that $\zeta^{4}(u)$ is associated with high-order fluctuating behavior
beyond gaussian approximation.

According to the minimal total dispersion, $\min(\Delta_{T}^{2})=2\lambda_{%
\Delta}(\kappa_{e})$, the system exhibits its largest energy fluctuations
when the curvature $\kappa_{e}$ reaches its minimum value $\kappa_{\min}$.
The character of the PT depends on the signature of the curvature $\kappa
_{\min}$. It is continuous for $\kappa _{\min }\geq 0$ ($q=2-4$), while
discontinuous for $\kappa _{\min }<0$ ($q>4$). The extended version of Wolff
algorithm is able to describe both continuous and discontinuous
temperature-driven PTs. Since 2D four-state Potts model exhibits the largest
fluctuating behavior near critical point, this particular case will be
considered to analyze the impact of the present methodology on the
efficiency of MC simulations.

For comparison purposes, the microcanonical quantities will be estimated
using the entropy $S(U)$ derived from Wang-Landau method \cite{WangLandau}.
To avoid statistical errors associated with numerical differentiation of the
entropy $S(U)$, we shall consider the formulae (\ref{EE}) of the point
statistical estimation. Statistical expectation values can be evaluated as
follows:
\begin{equation}  \label{sev.multi}
\left\langle a(U)\right\rangle=\frac{\sum_{i} a(U_{i})\exp\left[%
-\phi_{G}(U_{i})+S(U_{i})\right]}{\sum_{i} \exp\left[%
-\phi_{G}(U_{i})+S(U_{i})\right]},
\end{equation}
where $\phi_{G}(U)=\beta_{s}(U-U_{s})+\lambda_{s}(U-U_{s})^{2}/2N$. The
estimates of the entropy per site $s=S/N$ and the inverse temperature $\beta$
versus energy per site $u=U/N$ are shown in FIG.\ref{wanglandau.eps} for the
case of 2D four-state Potts model with $L=32$. As clearly illustrated here,
results obtained from a direct numerical differentiation of entropy $S(U)$
are strongly affected by the statistical errors associated with MC
calculations of energy histograms. Fortunately, the point statistical
estimation overcomes this difficulty providing a smoothly dependence for the
microcanonical caloric curve $\beta$ \emph{versus} $u$. Results from
Wang-Landau method are considered as a reference in FIG.\ref{thermo.comp.eps}%
, which illustrates microcanonical estimates derived from three canonical MC
algorithms and their extended versions. We have considered a variable number
of steps $M=4\times10^{4}\tau$ for each simulation run, with $\tau$ being
its decorrelation time. Dependencies associated with Wang-Landau method were
obtained from two simulation runs with $M=2\times 10^{7}$ and $M=1.1\times
10^{8}$ steps.

Results derived from the extended versions of canonical MC algorithms and Wang-Landau method exhibit a great agreement among them. Discrepancy among these MC methods is only observed for estimation of curvature curve $\kappa(u)$ near critical region (see in FIG.\ref{thermo.comp.eps}.b). This discrepancy was also observed in FIG.2 of Ref.\cite{vel-emc2}. In principle, extended canonical MC algorithms and Wang-Landau method should provide same microcanonical estimates when the number of steps $M$ is sufficiently large.
However, the entropy per site $s(u)$ obtained from Wang-Landau method is not sufficiently equilibrated to perform a more precise estimation of curvature curve $\kappa(u)=-\partial^{2}s(u)/\partial u^{2}$ near critical point. The convergence of results obtained from estimation formulae (\ref{EE}) is not uniform everywhere. Even using the optimal values for the seed parameters $U_{s},\beta_{s},\lambda_{s})$ of gaussian ensemble (\ref{GEns}), the largest fluctuating behavior is always observed near critical point. This fact evidences a particular advantage of extended canonical MC algorithms. These methods enables the study of a small energy region in a given simulation run. Thus, the number of steps $M$ of each run can be locally extended as large as needed to guarantee the convergence of microcanonical estimates. On the contrary, Wang-Landau method sweeps the whole energy range
in a single run. Although this feature is regarded as an advantage in many applications, this is not the case of calculations of partial derivatives $\partial^{n}s(u)/\partial u^{n}$. Statistical errors of entropy per site $s(u)$ are only reduced increasing the number of steps of Wang-Landau method for whole energy range. According to results shown in FIG.\ref{thermo.comp.eps}.b, there exist a certain converge of results of Wang-Landau method towards results of extended canonical MC methods when number of steps $M$ is increased from $2\times 10^{7}$ to $1.1\times 10^{8}$ \footnote{For implementing Wang-Landau multicanonical method described in Ref.\cite{mc3}, we have considered a minimum entry of 95\% of the mean value for histogram of energies visited. First simulation run with $M=2\times10^{7}$ steps was extended until parameter $f$ reaches the value $f=\exp(10^{-7})$. Second simulation run with $M=1.1\times10^{8}$ steps was extended until parameter $f$ reaches the value $f=\exp(10^{-8})$.}. However, the fully convergence requires much more calculations. This exigence contrasts with the high-performance of extended Wolff cluster algorithm. Using this last MC method, we have only employed a total of $M=7.3\times10^{6}$ steps, with an average of $M=2.2\times10^{5}$ steps for each calculated point.

\begin{figure}[t]
\begin{center}
\includegraphics[
width=3.5in ]{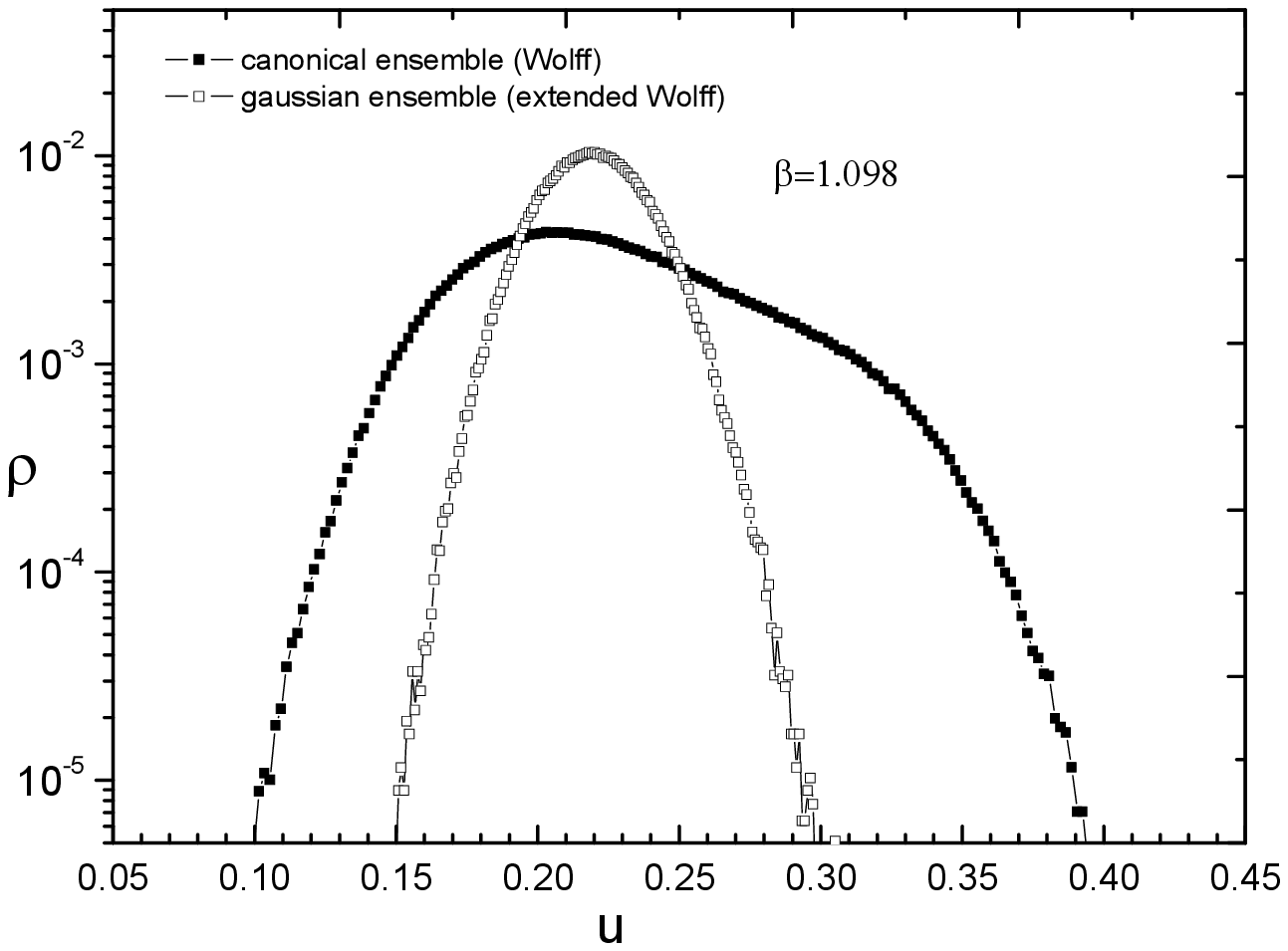}
\end{center}
\caption{Energy distributions associated with canonical ensemble and
gaussian ensemble with optimal values of the seed parameters $(U_{s},%
\beta_{s},\lambda_{s})$ near critical point. These results were
obtained from Wolff cluster algorithm and its extended version, respectively.
}
\label{systematic.deviation.eps}
\end{figure}

Microcanonical estimates derived from usual canonical MC algorithms undergo large systematic deviations. This behavior is not associated with a poorly equilibration of MC averages, but the large energy fluctuations experienced by this model system near critical point within canonical ensemble. Canonical ensemble is a particular case of gaussian ensemble with $\lambda_{s}=0$, so that, formulae (\ref{EE}) of the point statistical estimation are applicable to this ensemble whenever the associated energy distribution satisfies gaussian approximation. This requirement cannot be satisfied near critical point, which is is illustrated in FIG.\ref{systematic.deviation.eps}. We show here energy distributions near critical point obtained from MC simulations using both Wolff cluster algorithm and its extended version for $\beta\simeq 1.098$. Distribution obtained from usual Wolff cluster algorithm (canonical ensemble) cannot be described by a gaussian approximation. On the contrary, gaussian approximation is fully licit for distribution obtained from extended Wolff cluster algorithm, which considers a gaussian ensemble with optimal values of the seed parameters $(U_{s},\beta_{s},\lambda_{s})$. A way to reduce the incidence of finite size effects of microcanonical estimates derived from canonical ensemble is by considering higher-order correction terms in formulae (\ref{EE}). This exigency presupposes calculation of energy moments $\left\langle\delta U^{n}\right\rangle$ with $n>4$, which demands larger simulation runs to achieve their convergence.

The size dependencies of the decorrelation time $\tau$ and the efficiency factor $\eta$ at the critical point are shown in FIG.\ref%
{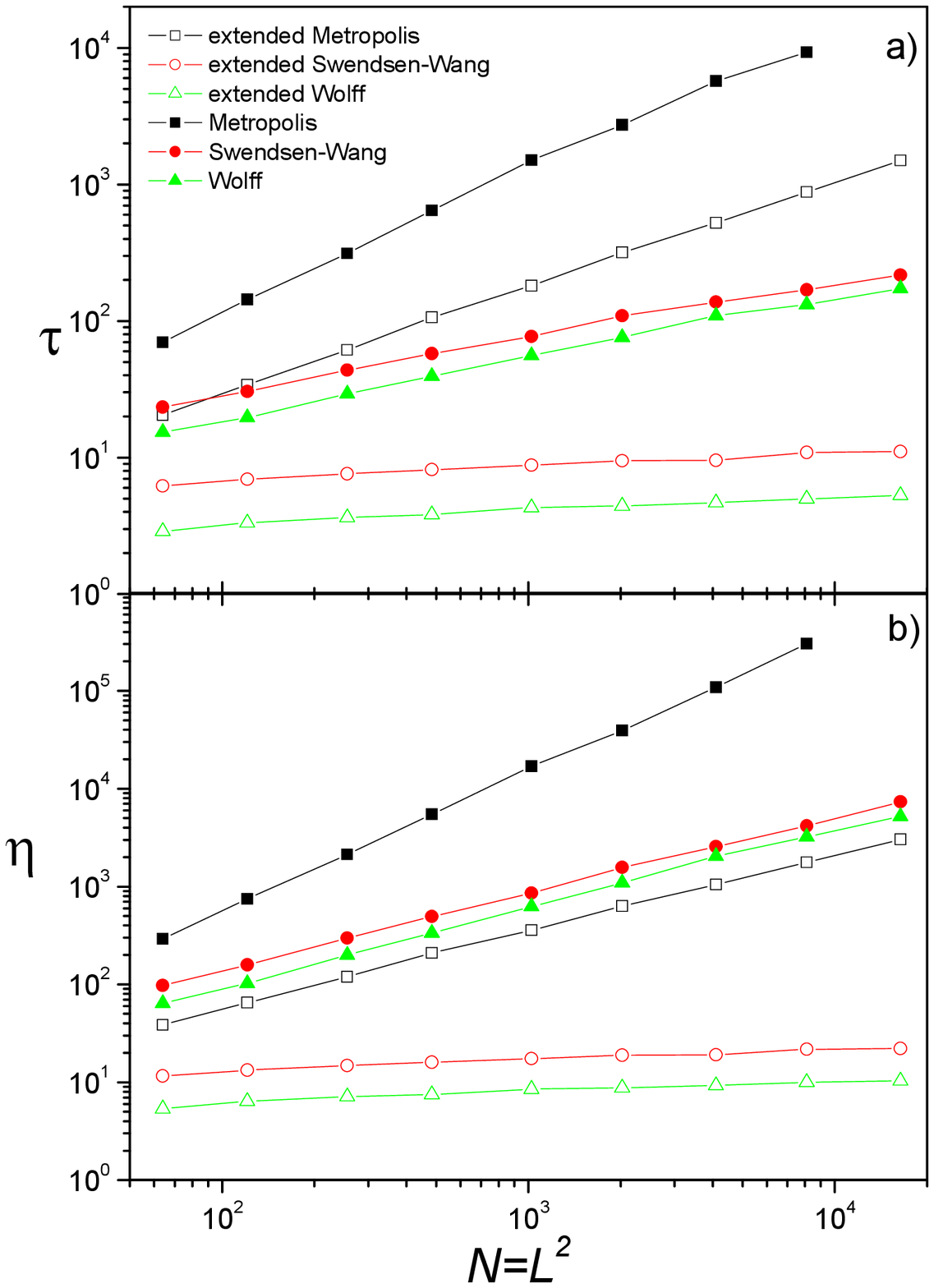} for canonical and extended versions of three different MC algorithms for lattice sizes $L$ ranging from $8$ to $128$. For all extended versions, size dependency of decorrelation time $\tau$ and the efficiency factor $\eta$ exhibit power-law behaviors $\tau(N)=C_{\tau}N^{\alpha_{\tau}}$ and $\eta(N)=C_{\eta}N^{\alpha_{\eta}}$ weaker than their canonical counterparts. For a better quantitative characterization, estimates of dynamic critical exponents $\alpha_{\tau}$ and $\alpha_{\eta}$ are shown in Table \ref{critical.exp}. The size dependency associated with Metropolis importance sample is reduced, but the improvement of its decorrelation time $\tau$ is less significant than the one achieved by cluster algorithms. Greater impact of the present methodology is manifested when the efficiency is described in terms of the efficiency factor $\eta$. Precisely, the efficiency factor $\eta$ determines the number of iterations needed to achieve the convergence of the microcanonical caloric curve $\beta(u)$. All extended MC algorithms exhibit a better efficiency factor $\eta$ than their original canonical counterparts. Extended version of Metropolis importance sample, in particular, is slightly more efficient than canonical Swendsen-Wang and Wolff cluster algorithms. The exponents for extended cluster algorithms near critical point $\alpha_{\eta}\simeq 0.1$, which are very similar to the typical values of systems that undergo temperature-driven discontinuous PT. Dynamic critical exponents $\alpha_{\tau}$ and $\alpha_{\eta}$ are practically the same for extended canonical MC algorithms. On the contrary, dynamic critical exponents of canonical MC algorithms exhibit a constant difference $\delta=\alpha_{\eta}-\alpha_{\tau}\simeq 0.36$ that is directly associated with the incidence of size effects in the total dispersion $\Delta^{2}_{T}$.

\begin{table}[tp]
\centering
\begin{tabular}{ccc}
\hline\hline
\textbf{MC method} & $\alpha _{\tau }$ & $\alpha _{\eta }$ \\ \hline\hline
\multicolumn{1}{c}{\small Metropolis} & $1.06\pm 0.01$ & $1.42\pm 0.01$ \\
\hline
\multicolumn{1}{c}{\small extended Metropolis} & $0.777\pm 0.006$ & $%
0.790\pm 0.008$ \\ \hline
\multicolumn{1}{c}{\small Swendsen-Wang} & $0.432\pm 0.007$ & $0.792\pm
0.008 $ \\ \hline
\multicolumn{1}{c}{\small extended Swendsen-Wang} & $0.098\pm 0.004$ & $%
0.117\pm 0.004$ \\ \hline
\multicolumn{1}{c}{\small Wolff} & $0.474\pm 0.005$ & $0.833\pm 0.007$ \\
\hline
\multicolumn{1}{c}{\small extended Wolff} & $0.094\pm 0.006$ & $0.103\pm
0.006$ \\ \hline\hline
\end{tabular}%
\caption{Dynamic critical exponents $\alpha _{\tau }$ and $%
\alpha _{\eta }$ associated with the size dependencies of
decorrelation time $\protect\tau $ and efficiency factor $\eta $
shown in FIG.\protect\ref{size.efficiency.eps}.}
\label{critical.exp}
\end{table}

\section{Final remarks}

Methodology proposed by Velazquez and Curilef \cite{vel-emc1,vel-emc2} leads
to a significant improvement of the efficiency of MC simulations in presence
of any type of temperature-driven phase transitions. Although extended
canonical cluster algorithms exhibit the highest efficiencies, a local MC
method as extended Metropolis importance sample has a universal
applicability and a very good efficiency. For the particular case of 2D
four-state Potts model, this extended local MC methods exhibits an
efficiency comparable to canonical cluster algorithms of Swendsen-Wang and
Wolff. Consequently, this extended local MC algorithm can be specially
useful in MC simulations of systems whose canonical cluster algorithms are
still unavailable in the literature.

Before to end this section, let us refer to some open problems. Firstly, the
present methodology should be extended to those MC algorithms based on
Boltzmann-Gibbs distributions \cite{Reichl}. An important antecedent of this
problem was considered by Velazquez and Curilef in Ref.\cite{vel-geft},
where general equilibrium fluctuation theorem (\ref{fdr}) was generalized
for the case of many thermodynamic variables. However, some relevant
developments still missing, as example, the extending of formulae (\ref{EE})
for the point statistical estimation. On the other hand, the present
methodology can be combined with re-weighting techniques such as
multi-histograms method to improve statistics \cite{mc3}, which can provide
a better estimation for the higher-order derivatives of the entropy $S(U)$.

\begin{figure}[t]
\begin{center}
\includegraphics[width=3.5in ]{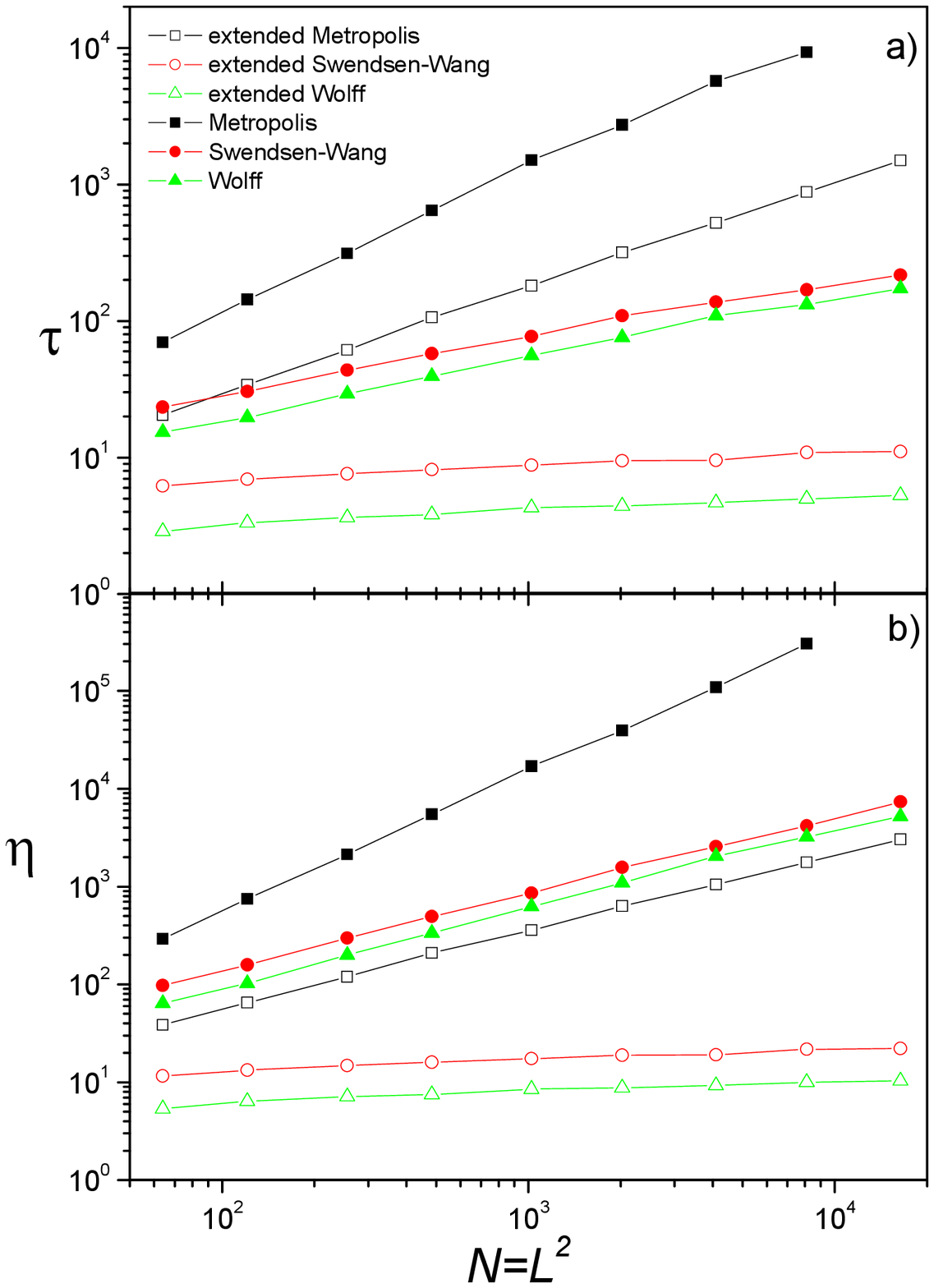}
\end{center}
\caption{(Color online). Size dependence of decorrelation time $\tau $ and
efficiency factor $\eta $ for three different canonical MC
algorithms and their extended versions at the critical point of 2D
four-state Potts model.}
\label{size.efficiency.eps}
\end{figure}

\section*{Acknowledgement}

L Velazquez thanks the financial support of CONICyT/Programa Bicentenario de Ciencia y Tecnolog\'{\i}a PSD 65 (Chilean agency).

\end{document}